\documentclass[sigconf, natbib]{acmart}
\usepackage{marvosym}

\usepackage{enumitem}
\usepackage{xcolor}
\usepackage{adjustbox}
\usepackage{bbding}
\usepackage{multirow}
\usepackage{multicol}
\usepackage{xspace}
\usepackage{bm}
\usepackage{caption}
\usepackage{subcaption}
\usepackage{makecell}
\definecolor{objective}{RGB}{159,221,246}
\definecolor{preference}{RGB}{255,210,208}
\definecolor{intention}{RGB}{190,254,204}
\newcommand{\paratitle}[1]{\vspace{1.5ex}\noindent\textbf{#1}}
\newcommand{\eg}{\emph{e.g.,}\xspace}
\newcommand{\ie}{\emph{i.e.,}\xspace}

\newcommand{\ignore}[1]{}

\AtBeginDocument{%
  \providecommand\BibTeX{{%
    \normalfont B\kern-0.5em{\scshape i\kern-0.25em b}\kern-0.8em\TeX}}}

\setcopyright{acmcopyright}
\copyrightyear{2022}
\acmYear{2022}
\acmDOI{XXXXXXX.XXXXXXX}

\acmConference[Conference acronym 'XX]{Make sure to enter the correct
  conference title from your rights confirmation email}{August 03--05,
  2022}{Woodstock, NY}

\acmPrice{15.00}
\acmISBN{978-1-4503-XXXX-X/18/06}

\begin{document}

\title{Recommendation as Instruction Following: A Large Language Model Empowered Recommendation Approach} 

\author{Junjie Zhang\textsuperscript{\textmd{1}}, Ruobing Xie\textsuperscript{\textmd{2}}, Yupeng Hou\textsuperscript{\textmd{1}}, Wayne Xin Zhao\textsuperscript{\textmd{1},\Letter},}
\author{Leyu Lin\textsuperscript{\textmd{2}}, and Ji-Rong Wen\textsuperscript{\textmd{1}}}

    \email{junjie.zhang@ruc.edu.cn,  batmanfly@gmail.com} 
    \affiliation{%
  \institution{\textsuperscript{\textmd{1}}Gaoling School of Artificial Intelligence, Renmin University of China}
  \institution{\textsuperscript{\textmd{2}}WeChat, Tencent}
   \country{China}
   }
   
    \thanks{\Letter\ Corresponding author.}

\renewcommand{\authors}{Junjie Zhang, Ruobing Xie, Yupeng Hou, Wayne Xin Zhao, Leyu Lin and Ji-Rong Wen}
\renewcommand{\shortauthors}{Zhang, et al.}

\begin{abstract}

In the past decades,  recommender systems have attracted much attention
in both research and industry communities, and a large number of studies have been devoted to developing effective recommendation models. Basically speaking,  these models mainly learn the underlying user preference from historical behavior data (typically in the forms of item IDs), and then estimate the  user-item matching relationships for recommendations.  

Inspired by the recent progress on large language models~(LLMs), we take a different approach to developing the recommendation models, considering recommendation as \emph{instruction following} by LLMs. The key idea is that the preferences or needs of a user can be expressed in natural language descriptions (called \emph{instructions}), so that LLMs can understand and further execute the instruction for fulfilling the recommendation task. 
Instead of using public APIs of LLMs, we  instruction tune an open-source LLM (3B Flan-T5-XL), in order to better adapt LLMs to recommender systems. For this purpose, we first design a general instruction format for describing the  preference, intention,  task form and context of a user in natural language.  
{Then we manually design 39 instruction templates and automatically generate a large amount of user-personalized instruction data (252K instructions) with varying types of preferences and intentions.}
To demonstrate the effectiveness of our approach, we instantiate the instruction templates into several widely-studied recommendation (or search) tasks, and conduct extensive experiments on these tasks with real-world datasets.  Experiment results show that the proposed approach can outperform several  competitive baselines, including the powerful GPT-3.5, on these evaluation tasks.  
Our approach sheds light on developing more user-friendly recommender systems, in which users can freely communicate with the system and obtain more accurate recommendations via natural language instructions.

\end{abstract}


\maketitle
\section{Introduction}
Nowadays,  recommendation systems have been widely deployed in various application platforms, which aim to satisfy user's needs and promote the use (or sale) of available resources. 
As the early approaches, collaborative filtering algorithms~\cite{sarwar2001item_collaborative,linden2003amazon_item_collaborative} were proposed by recommending items based on  similar tastes (\emph{user side}) or similar characteristics (\emph{item side}). Subsequently, matrix factorization~\cite{koren2009matrix_factorization} and neural networks~\cite{gru4rec,SASRec} were adopted  to develop the recommendation models,  which can capture more complex user preferences and learn more accurate user-item  relationships. 
Despite the huge progress, existing recommendation algorithms are mainly trained by fitting the user-item interaction data, lacking a generalization ability to unseen settings (\eg new users or new tasks). Besides, users are passively involved in conventional recommendation algorithms,  unable to explicitly express their real needs in a flexible form.

\begin{figure}[h]
	\centering
	\includegraphics[width=1.0\linewidth]{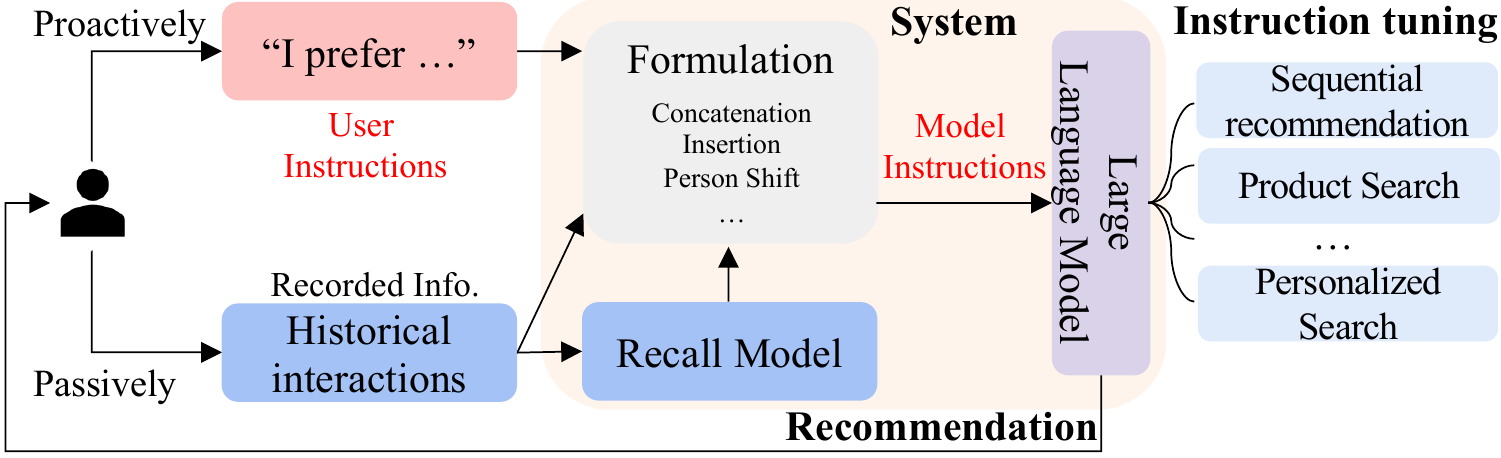}
	\caption{A framework of our proposed InstructRec.}
	\label{fig:illustration}
\end{figure}

Recently, pre-trained large language models~(LLM)~\cite{zhao2023survey_llm,zhang2022opt,touvron2023llama} (\eg T5~\cite{raffel2020exploring_t5} and GPT-3~\cite{gpt3}) have shown remarkable abilities on a variety of natural language tasks, which also shed lights on developing more general and effective recommender systems~\cite{hou2023learning_vq_rec,UniSRec,cui2022m6_rec,daiUncoveringChatGPTCapabilities,RLP,asai2022tart_task_aware_retrieval_instructions}. 
For example, it has been shown that  language models can improve the transferability of recommender systems~\cite{ZESRec,UniSRec,hou2023learning_vq_rec}, and also enhance the user-system interaction~\cite{RLP,cui2022m6_rec,wang2022towards_unicrs}.  
However, language models are built on natural language text, and they are not directly suitable for recommender systems, which need to deal with behavior data instead of text data.  
To bridge the gap between language models and recommender systems, a promising approach is to consider \emph{behavior modeling} as \emph{language modeling}~\cite{liu2023pre_plm_for_recommendation_survey,gaoChatRECInteractiveExplainable,cui2022m6_rec,RLP}: it formats the recommendation tasks in the form of natural language text and then treat them as a language processing task. In this approach, two fundamental problems are key to the success of item recommendations:  

\begin{enumerate}
    \item \emph{How to verbalize the recommendation tasks?} In general, a successful recommendation relies on the accurate understanding of the underlying user needs. It is important to design a suitable form to comprehensively describe the recommendation task in natural language, which should contain various kinds of useful information that reveal user's needs, including  the interaction history, user preference or intention, and other personalized factors.     
    \item \emph{How to adapt LLMs for  recommendation?}  Although LLMs are capable of modeling and generating the natural  language, it generally has difficulty in solving complex specialized tasks~\cite{fu2023specializing,liuChatGPTGoodRecommender2023} despite the generality. It requires specific tuning strategies that adjust the skills of LLMs towards the recommendation tasks, so as to bridge the gap and accurately understand user's behavior data.    
\end{enumerate}

In this paper, we would like to advance one step further in LLM-empowered recommender systems. Specially, we take a user-centric perspective, and consider the recommendation task as \emph{instruction following/execution} by LLMs. In this new recommendation paradigm, a user can freely express the specific needs via instruction texts, while the recommender system will follow the user's instructions for making accurate recommendations based on powerful LLMs. We make two major technical contributions: 

\begin{enumerate}
    \item \emph{A formal introduction  of instructions for recommendation}. We formally introduce the concept of \emph{instruction} in recommender systems. Despite the similar efforts in prior studies~\cite{RLP,asai2022tart_task_aware_retrieval_instructions}, we discuss the expression of user needs in the context of LLMs, and treat recommendation as a specific task approached by instruction following. Further, we present a detailed design for composing the instruction and instantiate it for solving different recommendation tasks with specific user needs or preferences. 
    \item \emph{An instruction tuned LLM for recommendation}. To adapt LLMs for recommender systems, we further employ instruction tuning to optimize  LLMs  according to the recommendation objective. For this purpose, we automatically generate a large number of recommendation-oriented instruction data for tuning the LLMs. Instruction tuning is key to elicit the abilities of LLMs for the recommendation tasks, and it further enables the generalization of LLMs on diverse user needs or requests.  
\end{enumerate}

To this end, we propose an instruction tuning approach for recommender systems, named \textbf{InstructRec}, which is a new recommendation paradigm that allows users to freely express their specific needs in natural language. To develop the recommender, our base model is built on the 3B Flan-T5-XL~\cite{chung2022scaling_flan_t5}\footnote{There is no consensus on the minimum size that discriminates between small-sized language models and LLMs. Our approach can be naturally extended to a larger language model. We refer to the readers to the survey~\cite{zhao2023survey_llm}  for a comprehensive review of LLMs. }. As the key to our approach, it is important to design the instruction format and construct the instruction data. Firstly, we formulate the instruction format by integrating four major parts, including preference, intention, task form and context.  As we will see, such an instruction form can cover a variety of specific user needs, and is particularly suitable for formatting different recommendation tasks.  {Secondly, we manually design 39 coarse-grained instruction templates to cover interaction scenarios,  and automatically generate 252K fine-grained user personalized instruction data with varying types of user preferences and intentions.} In order to generate high-quality  instruction data, we employ GTP-3.5 to generate the user preferences or intentions based on real historical data (\eg historical interaction or review texts)~\cite{wang2022self_instruct}. Further, we propose a series of useful strategies to increase the diversity of instructions, since it is important to improve the performance of instruction tuning~\cite{chung2022scaling_flan_t5, wei2021finetuned_flan}.    
 By tuning the LLM with these recommendation-oriented instruction data, the base model can be well adapted to recommender systems, and learn to follow the user's instructions for fulfilling the corresponding recommendation tasks.  
 Although we only consider single-turn instruction following, it would be promising to extend the current approach to a multi-turn conversation scenario. 
Such a task specialization approach of LLMs can enhance the overall generalization ability of the recommendation models and improve the user experiences in  recommender systems.

To evaluate the proposed approach, we conduct extensive experiments by constructing a wide range of interaction scenarios from real-world datasets. The proposed approach achieves promising performance compared to several competitive  baselines. The experimental results also demonstrate that InstructRec can effectively improve the LLM’s ability of accommodating the diverse user needs. Moreover, the performance on the held-out instructions and domains also verifies that our approach has a better generalization ability. The main contributions of this work are concluded as follows: 

\begin{itemize}
\item We cast recommendation as instruction following by LLMs, and introduce  a new recommendation paradigm that allows users to freely express their diverse information needs in recommendation. 

\item We design a flexible, general instruction form and automatically generate a large number of high-quality instruction data. Further, we fine-tune a 3B language model specially for recommender systems.  

\item Extensive experiments demonstrate the effectiveness and generalization ability of our approach on various task scenarios in recommender systems.
\end{itemize}

\section{Methodology}

In this section, we present the proposed  instruction  tuning approach for recommender systems, named InstructRec. Our approach allows users to freely express their information needs in  natural language instructions when interacting with the recommender system. To develop our approach, we first design a specific instruction format for recommendation, and formally introduce three key aspects (\ie  preference, intention, and task form) in instructions (Section~\ref{sec:1_instruction_format}).  Next, we introduce how to construct the instruction data according to various instantiations with different preference, intention, and task form  assisted by LLM, along with several strategies to increase the diversity of our instructions (Section~\ref{sec:2_instruction_generation}). Finally, we discuss how to fine-tune the base LLM with these generated instruction data for more user-centric recommendation (Section~\ref{sec:3_instruction_tuning}).

\subsection{Instruction Format for Recommendation}
\label{sec:1_instruction_format}
To enable a LLM with the ability to perform personalized recommendation tasks for users with instructions, the first step is to design a suitable instruction format, which aims to effectively reveal user's vague/specific intention, provide user's implicit/explicit preference, and clarify the task settings.  
In this part, we propose a unified instruction format containing three key aspects for recommendation, and then instantiate it for different interaction scenarios.

\begin{table*}[htbp]
  \centering
  \caption{Example instructions with various types of user \colorbox{preference}{preferences}, \colorbox{intention}{intentions}, and \colorbox{objective}{task forms}. To enhance the readability, we make some modifications to the original instructions that are used in our experiments.}
    \resizebox{\linewidth}{!}{\begin{tabular}{lll}
    \toprule
    \textbf{Instantiation} &\textbf{Model Instructions} \\
    \midrule
    $\langle P_1, I_0, T_0\rangle$ & The user has purchased these items: \colorbox{preference}{<historical interactions>}. Based on this information, \colorbox{objective}{is it likely} that the user will interact with \colorbox{objective}{<target item>} next? \\
    $\langle P_2, I_0, T_3\rangle$ &  
    You are a search engine and you meet a user's query: \colorbox{preference}{<explicit preference>}. Please respond to this user by selecting items from the candidates: \colorbox{objective}{<candidate items>.} \\
    $\langle P_0, I_1, T_2\rangle$ &  As a recommender system, your task is to recommend an item that is related to the user's \colorbox{intention}{<vague intention>}. Please \colorbox{objective}{provide your recommendation.}  \\
    $\langle P_0, I_2, T_2\rangle$ &  Suppose you are a search engine, now the user search that \colorbox{intention}{<specific Intention>}, can you \colorbox{objective}{generate the item} to respond to user's query?  \\
    $\langle P_1, P_2, T_2\rangle$ &  Here is the historical interactions of a user: \colorbox{preference}{<historical interactions>}. His preferences are as follows: \colorbox{preference}{<explicit preference>}. Please \colorbox{objective}{provide recommendations}.\\

    $\langle P_1, I_1, T_2\rangle$ &  The user has interacted with the following \colorbox{preference}{<historical interactions>}. Now the user search for \colorbox{intention}{<vague intention>}, please \colorbox{objective}{generate products} that match his intent.  \\
    $\langle P_1, I_2, T_2\rangle$ &  The user has recently purchased the following \colorbox{preference}{<historical items>.} The user has expressed a desire for \colorbox{intention}{<specific intention>.} Please \colorbox{objective}{provide recommendations.}  \\
    \bottomrule

    \end{tabular}}%
  \label{tab:example_instructions}%
\end{table*}%

\subsubsection{Key Aspects in Instructions}
\begin{figure}[t]
	\centering
	\includegraphics[width=1.0\linewidth]{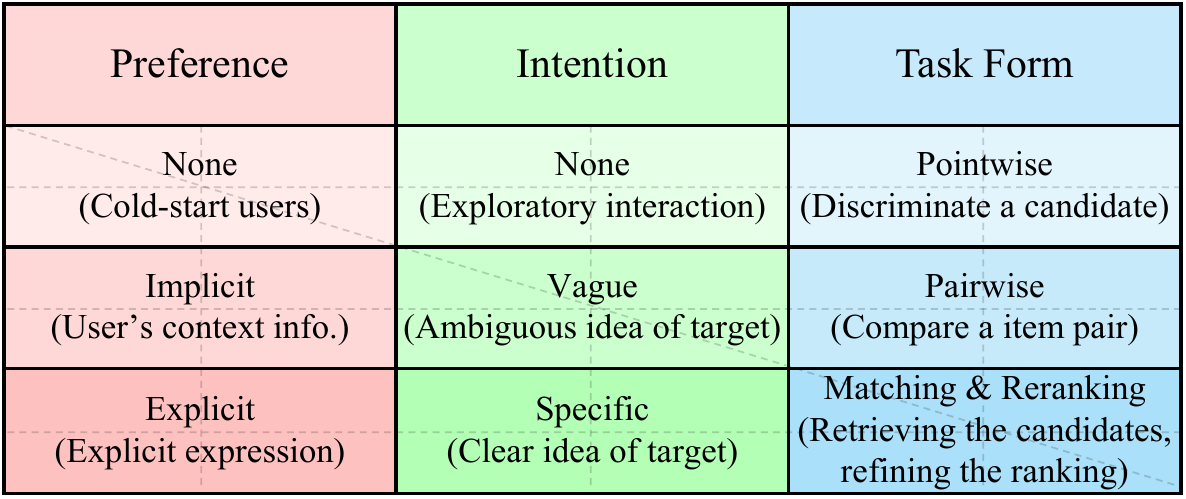}
	\caption{Various types of key aspects in instructions.}
	\label{fig:various_types_of_aspects}
\end{figure}

In order to design a flexible and expandable instruction format, we mainly consider three key aspects that are related to the expressions of user's needs, namely \emph{preference}, \emph{intention} and \emph{task form}.   
We present an illustration of the instruction format in Figure~\ref{fig:various_types_of_aspects}. 
In what follows, we introduce three aspects with their representative types in detail.

\paratitle{Preference~(P).} Preference refers to user's personalized tastes towards 
item attributes or characteristics. In our instruction format, it aims to capture inherent and long-term user preferences.  Based on the degree of personalization, user preferences can be categorized into the following three types:

$\bullet$ \emph{None $(P_0)$}. In this situation, the recommender system cannot obtain available information about user preferences or profiles, \eg cold-start  scenarios or under privacy concerns, which leads to non-personalized recommendations. 

$\bullet$ \emph{Implicit preference $(P_1)$}.
It refers that context information about a user (\eg user's profiles or historical interaction records) is available, but without explicit exposure of the underlying preference over items. In existing literature, implicit interaction behaviors are often called \emph{implicit feedback}~\cite{kelly2003implicit_feedback,hu2008collaborative_implicit_feedback}, which is easier to collect but may have noises. When formatting historical interaction records, we do not directly use item IDs to represent items as in P5~\cite{RLP}, but instead use item titles for formatting items as texts.   

$\bullet$ \emph{Explicit preference $(P_2)$}. In addition to the implicit preference, users may also directly reveal their preference in some cases. Such explicit feedback is more accurate but difficult to obtain. Here, we mainly consider the explicit expression of user preference in natural language text (\eg user reviews). We do not directly consider explicit interaction records (\eg ratings or likes), though they can be easily verbalized in our form.

\paratitle{Intention~(I).} Compared to long-term preferences, users' intentions refer to their more immediate demands for certain types of items, which may be different from user long-term preferences.  Inspired by Su et al.~\cite{su2018user_intent_in_query}, we categorize user intentions into three types according to the varying degree of clarity. 

$\bullet$ \emph{None $(I_0)$}. During the interaction, users may lack a clear goal for their next interactions and expect to discover potential interests through the recommender system and exploratory interactions.

$\bullet$ \emph{Vague intention~$(I_1)$}. In this case, users exhibit vague cognition about their needs, such as the descriptions of the desired characteristics of items, while can not clearly identify a specific item (\eg ``\emph{some gifts for my son}'').

$\bullet$ \emph{Specific intention~$(I_2)$}. In this case, users have clear needs, requiring the system to prioritize their requests and provide suitable items that satisfy the specific needs (\eg ``\emph{blue, cheap, iPhone13}'').

\paratitle{Task Form~(T).} In addition to user preferences and intentions, another  crucial aspect in formulating the instructions is the specified task form for execution by LLM. Similar to \cite{daiUncoveringChatGPTCapabilities}, we consider three  potential  forms for the recommendation tasks: 

$\bullet$ \emph{Pointwise recommendation~$(T_0)$}. The LLM is instructed to examine   whether a candidate item is appropriate for the user, where the matching relationship is determined based on user's information needs and item characteristics.

$\bullet$ \emph{Pairwise recommendation~$(T_1)$}. In this case, the comparison between a pair of items is invoked and the LLM is instructed to select the more suitable item from this pair.

{$\bullet$ \emph{Matching~$(T_2)$}. By acting as a matching (i.e., candidate generation) module, the LLM generates the potential candidates from the overall item corpus. These candidates are expected to be high-quality resources, and have a good coverage of the target items. }

$\bullet$ \emph{Reranking~$(T_3)$}. Unlike $T_2$ that instructs LLM to act as a matching module, the LLM is employed as a reranker in this case.  That is, the LLM is instructed to rerank the retrieved candidate items, enabling more accurate recommendations.

Besides the above three parts, we can add other useful context information (\eg time and place) about the user's situation, called \emph{context}.  
Overall, we can flexibly  integrate these four parts in the form of natural language text.

\subsubsection{Instantiation for Various Interaction Scenarios} 
\label{sec:intearction-scenarios}
In this part, we discuss how to instantiate the above instruction format for different real-world interaction scenarios. We present a list of examples in Table~\ref{tab:example_instructions} about instruction instantiation. Next, we introduce several representative instantiations. 

$\bullet$  $\langle P_1/P_2, I_0, T_3\rangle$.  In this case, we combine $P_1$ or $P_2$ with $I_0$ (without intention),  focusing on user preferences. When performings these instructions, LLMs act as \emph{a traditional recommender}. We can also combine both kinds of preferences, thereby instructing LLMs to perform recommendation with personalized prompts ($P_2$)~\cite{SASRec,wu2022personalized_prompt}.

$\bullet$  $\langle P_0, I_1/I_2, T_3\rangle$. We combine $P_0$ with $I_1$ or $I_2$ to instantiate the second type of instructions. Specially, LLMs act as \emph{a traditional retriever} to process users' vague or specific queries~\cite{huang2013learning_dssm}. 

$\bullet$ $\langle P_1/P_2, I_1/I_2, T_3\rangle$.  We take a combination between user preferences $(P_1/P_2)$ and intentions ($I_1/I_2$), aiming to rerank the candidate items.  Such a task can be also considered as \emph{personalized search}, where user preferences and intentions (queries) are available for satisfying user's needs~\cite{ai2017learning_hem,bi2020transformer_tem}. 

With this instruction format, we can instantiate different interaction scenarios. Note that there is no clear boundary between \emph{recommendation} and \emph{search} in our setting, since both can be formulated in a similar way (see the  discussion for $\langle P_1/P_2, I_1/I_2, T_3\rangle$). In the above, we mainly discuss the task form with $T_3$, since LLMs are currently more suitable for the reranking stage, due to the high cost of inference. In contrast, we can use other task forms such as $T_2$ for training, which is a more difficult task setting. Improving the diversity of instructions is verified to be effective in our evaluation.

\subsection{Instruction Generation}
\label{sec:2_instruction_generation}
According to the introduced instruction format in Section~\ref{sec:1_instruction_format}, we next generate the instruction data by 
simulating user preferences and intentions based on available interaction data. However, it is difficult to obtain a large amount of data that explicitly reveal real preferences or intentions of users.  {Recently, 
some efforts have attempted the automatic prompting strategies (\eg self-instruct~\cite{wang2022self_instruct}), which  generates high-quality instructions by prompting an instruction-tuned LLM (called \emph{teacher-LLM}). 
Motivate by this, we employ a strong teacher-LLM (\ie GPT-3.5) to generate such personalized information for each user by prompting with her/his historical interactions and reviews, which conveys rich information about preferences and intentions.} In what follows, we first present the pipeline of instruction generation, and then evaluate the quality of the generated instructions.

\subsubsection{Annotating the Aspects in Instructions.}

In order to better format the expressions of user needs via natural language instructions, we first manually create coarse-grained templates, according to the instantiations for various interaction scenarios (see Section~\ref{sec:intearction-scenarios}). Subsequently, we further fill these templates with specific user preferences and intentions (referring to fine-grained user personalized instructions) that are extracted from interaction data or generated by the teacher-LLM. In what follows, we illustrate the construction process in detail.

\paratitle{Preference annotation.} We use different strategies to generate the user preferences, considering the varying degrees of personalization. For the \emph{implicit preference} $P_1$, we take the title  of an item as its representation and utilize user's historical interactions to fill in the instruction templates, which can be instantiated as ``\emph{The user has previously purchased the following items: \{\texttt{[MASK]}\}}''. While, for the \emph{explicit preference} $P_2$, since explicit descriptions of user preferences are usually not available in the dataset,  the teacher-LLM (\ie GPT-3.5) is employed  to act as the user and generate explicit expressions of preferences based on the historical interactions. GPT-3.5 shows a strong text understanding ability,  and it can generate very meaningful expressions about users' preferences or intentions based on the historical interaction behaviors.  
Here is an example depicting the generation of user preference by GPT-3.5:
\begin{quote}
    [Raw Behavior Sequence]: ``\emph{1. Resident Evil: Revelations 2 - PlayStation 4 $\rightarrow$ 2. Resident Evil 4 - PlayStation 4 Standard Edition.}''
    \newline
    [Generated Explicit Preference]: ``\emph{He prefers horror-based games with a strong narrative.}''
\end{quote}

\paratitle{Intention annotation.} Similarly, we can generate user intentions in different clarity degrees, including vague intention $I_1$ and specific intention $I_2$. To derive the \emph{vague intentions}, since reviews provide valuable evidence about user's personal tastes and the reason to make a specific  interaction, we consider extracting the intentions from target reviews (the one  associated with the target item). Specially, the teacher-LLM is employed to process these reviews and extract the intention. Here is an example depicting the extraction of user's  intention from review text: 
\begin{quote}
    [Raw Target Review]: ``\emph{My son loves ... of the game. I'm happy I bought this for him.}''
    \newline
    [Generated Vague Intention]: ``\emph{I enjoy buying games for my son that he enjoys.}''
\end{quote} 
{As for the \emph{specific intentions}, users sometimes have a clear demand for certain items when using recommender systems (\eg buy a video game on e-commerce platforms). As discussed in previous work~\cite{bi2020transformer_tem,ai2017learning_hem}, the category of a  target item provides an important evidence to reflect the user's  real intention.
Consequently, we extract user's specific intention from the  category information of the target items, by concatenating multiple associated category labels into an intention expression:} 
\begin{quote}
    [Generated Specific Intention]: ``\emph{Video Games, PC, Accessories, Gaming Mice.}''
\end{quote}
Compared with the vague intention extracted from user's historical interactions, the extracted  intention is more specific, and can directly reflect user's real intention, for it is  extracted from the information of target items.  

\paratitle{Task form annotation.} In this paper, we mainly focus on three types of task forms: $T_0$, $T_2$, and $T_3$. Specifically, for \emph{pointwise recommedation} $T_0$, We formulate the instruction as: ``\emph{Based on the <user related information>, is it likely that the user will interact with <target item> next?}'', and  the system should respond with ``Yes'' or ``No''.  
{For the task of matching $T_2$, the instruction is formulated as ``\emph{Predict the next possible item}''. While for the task of \emph{reranking} $T_3$, a list of candidates is  incorporated into the instruction: ``\emph{Select one item from the following <candidates>}''.} Although currently we  do  not include the pairwise recommendation $T_1$, it can be easily integrated into our proposed framework.

\subsubsection{Increasing the Diversity of Instructions.}
Recent studies~\cite{wei2021finetuned_flan,chung2022scaling_flan_t5} have demonstrated the effectiveness of increasing the quantity and diversity of instructions. Therefore, in order to further improve the performance of instruction tuning, we propose to use several strategies to increase the diversity of instruction data.  

\paratitle{Turn the task around.} This strategy refers to the  swap between the input and output of normal instructions~\cite{wei2021finetuned_flan}. In our case, we request LLM to not only recommend the appropriate items based on user preferences or intentions, but also infer their potential information needs based on the feedback of recommendation. Such tasks can help LLM understand the relationship between user behaviors and the underlying  information needs. An example of instruction and its reversed version is shown below:
\begin{quote}
    [Normal Instruction]: \emph{``The user searches that <Query>, can you generate the item to respond to his query?''}
    \newline
    [Swapped Instruction]: \emph{``The user wants to buy: <Target item>, but he doesn't know how to formulate the query, please help him generate it.''}
\end{quote}

\paratitle{Enforcing the relatedness between preference and intention.} It refers that the revealed short-term user intentions with long-term preferences should be highly related  in an instruction.    

\begin{quote}
    [Intention $\rightarrow$ Preference]: \emph{``The user has searched for <Query> and selected the product <Target item>. Based on his query and choice, please infer his historical interactions.''}
    \newline
    [Preference $\rightarrow$ Intention]: \emph{``The user has the following <historical interactions>, you can infer their preferences. Please make a prediction about the user's next query and the item he is likely to purchase, based on his preference.''}
\end{quote}

\paratitle{Chain-of-thought~(CoT) like reasoning.} This strategy adds additional explanations at 
intermediate reasoning steps, that enables LLM to perform complex reasoning tasks~\cite{wei2022chain_of_thought_cot}.  
To apply this idea to our tasks, it refers to
the reasoning process from user's implicit behaviors to explicit preferences or intentions that lead to the final recommendations: 

\begin{quote}
    [CoT Instruction]: \emph{``Given the <Historical Interactions> of the user, please infer his preference and recommend suitable items to him.''}
    \newline
    [Desired Responses]: \emph{``According to the user's historical interactions, we can infer his <preferences>. Finally, we recommend him <target item>.''}
\end{quote}

\subsubsection{Statistics and Quality  of Instruction Data.}

Based on the above steps, we first manually design 39 coarse-grained instruction templates (
see a complete list of instruction templates in the Appendix), covering a diverse range of different interaction scenarios. Then we generate a total of 252K fine-grained  user instructions, which describe user various types of preferences and intentions, with the help of the teacher-LLM (\ie GPT-3.5). The statistics of the generated instructions are summarized in Table~\ref{tab:instruct_statistics}.  
{To assess the quality of these fine-grained instructions, we randomly sample 100 instances from each of the instructions describing user preferences and intentions for evaluation.}
An annotator is asked to evaluate the appropriateness of the generated instructions based on their consistencies with the user data (\eg  historical interactions, target item and the associated review). The results are illustrated in Table~\ref{tab:quality_evalutation}. In most cases, the teacher-LLM can generate appropriate  instructions based on  user's specific information. 
Not limited to the original information, it can also produce more detailed instructions by leveraging the encoded world knowledge. 
Currently, only 69\% of the instructions accurately align with the user's current intention, and we speculate that it is affected by the contained noise in the review text. We will explore how to generate more accurate intentions as the future work.

\begin{table}[t]
  \centering
  \caption{Statistics of the annotated instructions.}
    \begin{tabular}{lr}
    \toprule
    \textbf{Statistic} &  \\
    \midrule
    \specialrule{0em}{0.5pt}{0.5pt}
    \midrule
    \# of fine-grained instructions & 252,730 \\
    \quad - \# of user-described preferences & 151,638 \\
     \quad  - \# of user intention in decision making & 101,092 \\
    ave. instruction length~(in words) & 23.5 \\
    \midrule
    \# of coarse-grained instructions & 39 \\
    \quad - \# of preferences related instructions & 17 \\
     \quad  - \# of intentions related instructions & 9 \\
    \quad   - \# of combined instructions & 13 \\
    ave. instruction length~(in words) & 41.4 \\
    \bottomrule
    \end{tabular}%
  \label{tab:instruct_statistics}%
\end{table}%
\begin{table}[t]
  \centering
  \caption{Quality evaluation of the generated fine-grained instructions by taking text-davinci-003 as teacher-LLM.}
    \begin{tabular}{ccc}
    \toprule
    \multicolumn{1}{c}{\textbf{Quality Review Question}} & \textbf{Preference} & \textbf{Intention} \\
    \midrule
    \specialrule{0em}{0.5pt}{0.5pt}
    \midrule
    \makecell{Is the instruction generated from \\ the user's related information?} & 93\%  & 90\% \\
    \midrule
    \makecell{Does the teacher-LLM provide \\ related world knowledge?} & 87\%  & 22\% \\
    \midrule
    \makecell{Does the instruction reflect \\ the user's preference/ intention?} & 88\%  & 69\% \\
    \midrule
    \makecell{Is the instruction related to \\ target item?} & 48\%  & 69\% \\
    \bottomrule
    \end{tabular}%
  \label{tab:quality_evalutation}%
\end{table}%

\subsection{ Instruction Tuning for Recommendations}
\label{sec:3_instruction_tuning}
Based on the above instruction format,
we then discuss how to instruction tune LLMs for recommender systems. We first present the backbone LLM, and then introduce the optimization and inference. 

\subsubsection{The Backbone LLM}
In this work, we aim to develop a recommender model  capable of following user-centric instructions that specify their needs.
Inspired by the recent progress on LLMs~\cite{zhao2023survey_llm, touvron2023llama, rae2021scaling_gopher,hoffmann2022training_Chinchilla}, our approach is based on the fact that LLMs exhibit strong abilities in following user's instructions to solve various tasks. 
Using LLMs as the recommender, we treat recommendation as a specific task for instruction following. It has been shown that instruction tuning  enables LLMs to  generalize to unseen tasks described in natural language instruction~\cite{wei2021finetuned_flan,ouyang2022training_instruct_tuning_align_user_intent}. Thus, such an approach can  be potentially extended to fulfill more diverse tasks in application systems. While, we limit our discussion to the recommendation task.  
Specially, we adopt the 3B Flan-T5-XL~\cite{chung2022scaling_flan_t5} as the backbone model. Since Flan-T5 has been fine-tuned based on T5~\cite{raffel2020exploring_t5} with a large amount of  instruction data, it has an excellent capacity to follow natural language instructions.   
However, the original instructions for tuning Flan-T5 are not specially designed for recommender systems, which  cannot effectively instruct the model to perform the recommendation task. 
To be specific, Flan-T5 is designed with an encoder-decoder architecture, supporting a maximum context length of 512 tokens. 
In our approach, we represent items using their associated texts (\eg titles), which might  result in an excessive input length when formatting users' behavioral sequences. In such cases, we have to truncate the input sequence and likely result in suboptimal performance. 
Researchers can alternatively utilize LLMs with a longer  context length, such as LLaMA~\cite{touvron2023llama}, to model long behavioral sequences.

\subsubsection{Training and Inference}
In this part, we discuss how to optimize our base LLM with the generated instructions and how to use it for solving the recommendation tasks. 
 
\paratitle{Optimization via Instruction Tuning.} With the generated instruction data, we can optimize the LLM via instruction tuning, which is essentially  a supervised fine-tuning way. Specifically, we first annotate the desired system responses (target output), according to different types of instructions. For example, when instructing the model to predict the next item, the target output is annotated as the target item. While, for CoT like instructions, the target output is annotated as the user's reasoning process for the specific  interaction. Since both the instruction and the target output can be formatted in natural language, we can unify the training into a unified sequence-to-sequence way. Formally,  we optimize the negative log-likelihood of the target output as follows:
\begin{align}
\label{eqn:loss}
    \mathcal{L}=\sum_{k=1}^{B}\sum_{j=1}^{|Y_k|} \log P\left(Y_{k,j} \mid Y_{k,<j}, I_{k}\right),
\end{align} 
{where $Y_k$ is the desired system responses for the $k$-th instance, $I_k$ is the instruction of the $k$-th instance, and $B$ is the batch size.} 

\paratitle{Inference.}
Via instruction tuning, the LLM has been trained to  follow the instructions in various interaction scenarios.  {In this part, we present the application of our approach in recommender systems. Considering the  computational efficiency and model capacity, we employ the LLM as a reranker to generate the final ranking for the candidates based on users' instructions. 
In real-world systems, users' needs are very diverse, and we expect that instructions can effectively capture different preferences or intentions by leveraging the generality of natural language. 
Specifically, when providing the recommendation services to users, the system will firstly  select appropriate coarse-grained instruction templates based on \emph{user instructions} (\ie the instructions that a user issues)  and other useful information (\eg  historical interactions). Then, we convert 
the original expressions into \emph{model instructions} by using the operations such as concatenation, insertion, and persona shift. 
Then, the recommender (\ie LLM) is requested to execute the model instruction that specifies user's needs. 
However, due to the inherently stochastic nature of the generation process in LLMs, there exists a potential risk of generating items outside of the candidate set, especially when using beam search to generate a list of candidate items. To circumvent this problem, similar to Eq.~\eqref{eqn:loss}, we directly feed the candidate items as input to the decoder of the model and calculate their likelihood for determining the final ranking for recommendations.}

\begin{table}[t]
  \centering
  \caption{Comparison between the proposed InstructRec and
two related studies.   ``IT'' denotes instruction tuning.}
    \begin{adjustbox}{width=1.0\columnwidth}
    \begin{tabular}{llccl}
    \toprule
    \textbf{Methods} & \textbf{Backbone} & \textbf{IT} & \textbf{Non-ID} & \textbf{Key point} \\
    \midrule
    M6-Rec & 300M M6 & \color[RGB]{191,0,64}{\XSolidBrush}     & \color[RGB]{0,128,128}{\CheckmarkBold}    & Representing user behavior as plain texts \\
    P5    & 220M T5-base & \color[RGB]{0,128,128}{\CheckmarkBold}     & \color[RGB]{191,0,64}{\XSolidBrush}     & Task description \\
    InstructRec & 3B Flan-T5-XL & \color[RGB]{0,128,128}{\CheckmarkBold}     & \color[RGB]{0,128,128}{\CheckmarkBold}     & Aligning  LLMs with  user needs \\
    \bottomrule
    \end{tabular}%
    \end{adjustbox}
  \label{tab:compare_llm4rec}%
\end{table}%

\subsection{Discussion}
In this part, we compare the proposed InstructRec with related methods, thereby highlighting the contributions of our approach.

\paratitle{Traditional methods}  such as SASRec~\cite{SASRec} and LightGCN~\cite{he2020lightgcn} typically rely on unique identifiers to represent users and items,  and construct specific preference  functions for recommendations. However, they often struggle to address the cold-start problem, as new users or items cannot be well represented.   Additionally, traditional recommendation approaches focus on  passive acceptance of recommendations by users,   potentially failing to capture the true preferences accurately. While user can actively request a search engine to retrieve relevant items~\cite{huang2013learning_dssm,shen2014learning_cdssm}, traditional search engines can only handle specific user queries due to their limited capacities. 
As a comparison, our model formulates the recommendation task using natural language and leverages universal knowledge in LLM to make the recommendations, having a better  generalization ability.  Furthermore, by fine-tuning on instructions with varying types of user preferences, intentions and task forms, our model can accommodate diverse user's needs effectively.

\paratitle{Existing applications of LLMs in recommender systems} such as P5~\cite{RLP} and M6-Rec~\cite{cui2022m6_rec} consider behavior modeling as language modeling, where recommendation tasks are formulated as natural language expressions. M6-Rec reduces the reliance on user behavioral data by incorporating and training on a wide range of contextual information via converting various kinds of context information into a sequential text format, thereby achieving strong generalization across different interaction behaviors. P5 formulates multiple recommendation tasks with instruction-based prompts and demonstrates strong performance on multiple tasks. However, they mainly focus on task-specific formulations, while neglect the exploration of aligning  LLMs with \emph{more diverse, detailed user  needs} in practical scenarios. In contrast, our approach designs and generates recommendation-oriented instructions containing different types of preferences, intentions, and task forms, which can enhance the personalization in various interaction scenarios.
The comparison of our approach and the two related studies is presented in Table~\ref{tab:compare_llm4rec}.

\section{Experiments}
In this section, we conduct experiments to evaluate the ability of our proposed \emph{InstructRec} in several aspects.

\subsection{Experimental Setup}

\subsubsection{Datasets}
We evaluate our model's performance on accommodating user-centric instructions with the ``Video Games'' subset of Amazon\footnote{https://nijianmo.github.io/amazon/index.html} dataset and its generalization to unseen data with ``CDs \& Vinyl'' subset of Amazon dataset respectively. Following previous work~\cite{UniSRec}, we filter unpopular users and items with fewer than five interactions for all datasets. Since Flan-T5 (our backbone) has a contextual limit of 512 tokens, we truncate the generated behavioral sequence with a maximum of 20 items. The statistics of our preprocessed datasets are summarized in Table~\ref{tab:datasets}.

\begin{table}[!hbpt]
\centering	
 \caption{Statistics of the datasets after preprocessing. ``Avg.$len$'' represents the average length of item sequences.}
	\label{tab:datasets}
    
		\resizebox{\columnwidth}{!}{\begin{tabular}{lrrrrr}
			\toprule[1pt]
			\textbf{Dataset}  & \textbf{\#Users} & \textbf{\#Items} & \textbf{\#Inters} & \textbf{\#Sparsity} & \textbf{Avg.$len$} \\	
			\midrule
            \textbf{Games} & 50,546 & 16,859 & 410,907 & 99.952\% & 8.13 \\
			\textbf{CDs} & 93,652 & 63,929 & 871,883 & 99.985\% &9.31 \\
			\bottomrule[1pt]
		\end{tabular}}
\end{table}

\subsubsection{Evaluation Metrics}
We adopt top-$K$ hit ratio (HR) and top-$K$ normalized discounted cumulative gain (NDCG) to evaluate the performance. In this paper, K is set to 1, 3 and 5. For interaction scenarios such as sequential recommendation and personalized search, following the previous work~\cite{S3Rec,UniSRec,bi2020transformer_tem}, we apply the \textsl{leave-one-out} strategy for evaluation. Concretely,  for each user interaction sequence, the last item and its related review are used as the test data, the item and review before the last interaction are used as the validation data, and the remaining interaction records are used for training. 
For interaction scenarios such as product search, we split all the items and their related queries into training (80\%), validation (10\%) and testing (10\%). In this setting, instances of validation and testing set are unseen in the training stage, which increases the difficulty of inference. We implement some baselines with a popular open-source recommendation library RecBole~\cite{zhao2021recbole_1,zhao2022recbole_recbole_2,xu2022recen_recbole_3}. Notably, for different types of coarse-grained instructions, we evaluate the results of instruction that obtains the best performance in the valid set. By taking the model as a reranker, we rank the ground-truth of each instance among nine randomly sampled negative items, for evaluation on the test set (we further evaluate on harder settings in Section \ref{sec.further_analyses}), and finally report the average score of all test instances.

\subsection{Overall Performance on Various User Information Needs}
To demonstrate the effectiveness of our model in accommodating diverse user instructions, we conduct in-depth experiments according to the instantiations of these instructions for various interaction
scenarios (see instantiations in Section~\ref{sec:intearction-scenarios}). In what follows, we introduce these experiments in different scenarios separately, including the baselines and the performance comparisons. Notably, in addition to introducing baselines that specialized in specific interaction scenarios, we further take text-davinci-003~\footnote{https://platform.openai.com/docs/models/overview}, which is one of the universal GPT-3.5 model, as a strong LLM baseline for comparison. Notably, due to the high cost of this model, we random sample 500 instances to evaluate its performance. Although this may inevitably introduce some randomness, we believe that the conclusions obtained in this setting are still referable for exploring the recommendation capabilities of a universal LLM. 

\subsubsection{Sequential Recommendation $\langle P_1, I_0, T_3 \rangle $.}
We first conduct an evaluation of our model's performance in the classical sequential recommendation task by formulating the instructions with user's implicit preference $P_1$ (\eg behavioral sequences). 

\paratitle{Baseline.}
We adopt \textbf{SASRec}~\cite{SASRec} and \textbf{BERT4Rec}~\cite{BERT4Rec} as our baselines in the scenario of sequential recommendation. SASRec is a representative sequential recommendation model that utilizes a transformer-encoder architecture, incorporating a multi-head self-attention mechanism to effectively capture sequential patterns in the data. BERT4Rec is also widely used, which incorporates bidirectional self-attention mechanisms, leveraging a cloze objective to effectively encode sequential data.

\begin{table}[htbp]
\setlength{\belowcaptionskip}{-2pt}
  \centering
  \caption{Performance on sequential recommendation.}
    \resizebox{\columnwidth}{!}{\begin{tabular}{l|ccccc}
    \toprule
    \multirow{2.5}{*}{\textbf{Methods}} & \multicolumn{5}{c}{\bm{$\langle P_1, I_0, T_3 \rangle $}} \\
\cmidrule{2-6}          & HR@1  & HR@3  & HR@5  & NDCG@3 & NDCG@5 \\
    \midrule
    BERT4Rec & 0.5747 & 0.8188 & 0.9083 & 0.7176 & 0.7546 \\
    SASRec & 0.6663 & 0.8741 & 0.9389 & 0.7887 & 0.8155 \\
    GPT-3.5 & 0.3640 & 0.6300 & 0.7300 & 0.5216 & 0.5623 \\
    \midrule
    InstructRec & \textbf{0.6947} & \textbf{0.8793} & \textbf{0.9429} & \textbf{0.8033} & \textbf{0.8295} \\
    Improv. & +4.26\% & +0.59\% & +0.43\% & +1.85\% & +1.72\% \\
    \bottomrule
    \end{tabular}}%
  \label{tab:eval_seq_rec}%
\end{table}%

\begin{table*}[htbp]
  \centering
  \caption{Performance on personalized search. We have evaluated on three types of queries built from $P_2$, $I_1$, and $I_2$.}
    \begin{adjustbox}{width=1.0\linewidth}
    \begin{tabular}{l|ccccc|ccccc|ccccc}
    \toprule
    \multirow{2.5}{*}{\textbf{Methods}} & \multicolumn{5}{c|}{\bm{$\langle P_1, P_2, T_3 \rangle $}}   & \multicolumn{5}{c|}{\bm{$\langle P_1, I_1, T_3 \rangle $}}   & \multicolumn{5}{c}{\bm{$\langle P_1, I_2, T_3 \rangle $}} \\
    \cmidrule(lr){2-6}\cmidrule(lr){7-11}\cmidrule(lr){12-16} & HR@1  & HR@3  & HR@5  & NDCG@3 & NDCG@5 & HR@1  & HR@3  & HR@5  & NDCG@3 & NDCG@5 & HR@1  & HR@3  & HR@5  & NDCG@3 & NDCG@5 \\
    \midrule
    TEM   & 0.5005 & 0.7368 & 0.8462 & 0.6383 & 0.6833 & 0.5723 & 0.7860 & 0.8712 & 0.6974 & 0.7325 & 0.8665 & 0.9149 & 0.9293 & 0.8954 & 0.9013 \\
    GPT-3.5 & 0.2740 & 0.5500 & 0.7000 & 0.4366 & 0.4979 & 0.3660 & 0.6360 & 0.7566 & 0.5256 & 0.5795 & 0.4580 & 0.7240 & 0.8100 & 0.6138 & 0.6499 \\
    \midrule
    InstructRec & \textbf{0.6959} & \textbf{0.8806} & \textbf{0.9452} & \textbf{0.8045} & \textbf{0.8312} & \textbf{0.8278} & \textbf{0.9444} & \textbf{0.9741} & \textbf{0.8971} & \textbf{0.9094} & \textbf{0.9269} & \textbf{0.9868} & \textbf{0.9951} & \textbf{0.9631} & \textbf{0.9665} \\
    Improv. & +39.04\% & +19.52\% & +11.70\% & +26.04\% & +21.64\% & +44.64\% & +20.15\% & +11.81\% & +28.63\% & +24.15\% & +6.97\% & +7.86\% & +7.08\% & +7.56\% & +7.23\% \\
    \bottomrule
    \end{tabular}%
    \end{adjustbox}
  \label{tab:eval_personalized_search}%
\end{table*}%

\paratitle{Performance comparison.}
For sequential recommendation, users do not actively express their information needs to the system, which poses certain limitations on our model's capabilities. Nevertheless, through finetuning on large amounts of behavioral data and leveraging the generalization of instructions, as demonstrated in Table~\ref{tab:eval_seq_rec}, our model outperforms other baselines. Furthermore, we find that GPT-3.5 does not exhibit satisfactory performance in this particular scenario. This can be attributed to the mismatch between the universal textual nature of LLM and the specificity of behavioral information in private domain. In contrast to natural language, user behavioral sequences are highly personalized and exhibit greater complexity, posing a challenge for universal LLMs to capture individual behavioral patterns. 
Therefore, in order to deploy LLMs in private domain recommender systems, a proper fine-tuning rather than simply relying on LLMs under the zero-shot setting may be crucial for capturing personalized user behaviors.

\subsubsection{Product Search $\langle P_0, I_2, T_3 \rangle$}
In this part, we evaluate the efficacy of our model on product search. A typical product search task takes a search query from users and then recommends items that are related to the query and will be clicked by the user. Here, the search query is simulated by user specific intention $I_2$ extracted from target items' metadata, such as target item categories.

\paratitle{Baseline.}
We take \textbf{DSSM}~\cite{huang2013learning_dssm} as our baseline. DSSM is a widely-verified retrieval model that employs a two-tower architecture, aiming to establish a semantic-level mapping between a given user query and relevant documents. In our implementation, we leverage BERT-mini~\cite{bert}, which is known for robust encoding capabilities, to represent and encode the query and document inputs.

\begin{table}[!htbp]
\vspace{-1pt}
  \centering
  \caption{Performance on product search.}
    \resizebox{\columnwidth}{!}{\begin{tabular}{l|ccccc}
    \toprule
    \multirow{2.5}{*}{\textbf{Methods}} & \multicolumn{5}{c}{\bm{$\langle P_0, I_1, T_3 \rangle$}} \\
\cmidrule{2-6}          & HR@1  & HR@3  & HR@5  & NDCG@3 & NDCG@5 \\
    \midrule
    DSSM  & 0.7279 & \textbf{0.9484} & \textbf{0.9899} & 0.8587 & 0.8760 \\
    GPT-3.5 & 0.6700 & 0.9140 & 0.9480 & 0.8177 & 0.8399 \\
    \midrule
    InstructRec & \textbf{0.8263} & 0.9411 & 0.9728 & \textbf{0.8944} & \textbf{0.9075} \\
    Improv. & +13.52\% & --    & --    & +4.16\% & +3.60\% \\
    \bottomrule
    \end{tabular}}%
  \label{tab:eval_product_search}%
\end{table}%

\paratitle{Performance comparison.}
For the evaluation of product search, wherein the instructions are relatively specific (simulated from item category), the traditional model performs well. Furthermore,  since the items in the test set would be unseen during the training phase, 
this evaluation necessitates a higher degree of generalization capacity. Therefore, as we can see in Table~\ref{tab:eval_product_search}, our model achieves superior or comparable performance in most cases (especially for top ranking metrics such as NDCG@1) due to the vast amount of general knowledge encoded in its parameters. This conclusion can also be supported by the good results of GPT-3.5 without tuning.

\subsubsection{Personalized search $\langle P_1, P_2/I_1/I_2, T_3 \rangle$.}
In this part, we evaluate the model's capacity in personalized search, which could be viewed as a natural combination of recommendation and search tasks. Notably, in the literature of personalized search~\cite{ai2017learning_hem,dou2007large_personalized_search_survey)}, there are various ways to inject personalization into traditional search engines, such as introducing user's historical requests or clicks in logs. In this paper, following previous work~\cite{bi2020transformer_tem,ai2017learning_hem}, we conduct experiments in the latter setting. Specifically, we leverage the historical behavior sequences ($P_1$) for the ``personalized'' part. For the ``search'' part, we comprehensively adopt the user explicit preference simulated by LLMs $(P_2)$, vague intention $(I_1)$, and specific intention $(I_2)$ as three types of queries respectively. 

\paratitle{Baseline.}
We introduce \textbf{TEM}~\cite{bi2020transformer_tem} as our baseline in this scenario. As a representative approach in personalized product search, TEM utilizes a transformer architecture to encode the sequences of query and user's behavioral sequence, thereby achieving dynamic control over the impact of personalization on the search results. 

\paratitle{Performance comparison.}
The performance evaluation of our model in accommodating personalized search is presented in Table~\ref{tab:eval_personalized_search}. We observe that:
(1) in general, our InstructRec outperforms other approaches by a large margin in almost all the cases. Specifically, when taking user specific intention (\ie item category) as instruction, despite the provision of additional supplemental information such as user behavioral data, GPT-3.5 exhibits worse performance in comparison to its effectiveness in conducting product search. This observation emphasizes the huge gap between user behavioral patterns of private domain data and the universal semantic knowledge encoded within LLMs. While leveraging the technique of instruction tuning, our model, which is also built upon a universal LLM, effectively bridges the gap and aligns itself with user's personalized behaviors within recommender systems.
(2) In scenarios where user instructions exhibit relative ambiguity, such as LLM-generated explicit preference and vague intention, traditional models tend to perform poorly. This can be attributed to the incapacity of traditional models to capture the underlying vague information needs of users conveyed in these instructions. 
(3) Furthermore, although explicit user preferences are typically simulated by analyzing historical interactions and most of them are reflective of user's real preferences, there are still observed mismatches between user's long-term mainstream preferences ($P_2$) and the current intentions of target items in the test set (see the quality evaluation of instructions in Table~\ref{tab:quality_evalutation}). Therefore, the results of $\langle P_1, P_2, T_3 \rangle $ are much worse than those of $\langle P_1, I_2, T_3 \rangle $ whose queries are more related to user's real intentions. Nevertheless, the introduction of instruction tuning empowers our model with reasoning and generalization capabilities, thereby enabling it to perform complex interaction scenarios involving ambiguous instructions.

To sum up, through instruction tuning, our model effectively integrates user behaviors in private domain with universal knowledge. Consequently, it achieves the best performance in almost all classical practical tasks including sequential recommendation, product search, and personalized search, regardless of whether the user's preferences are implicit or explicit, and whether the intention is vague or specific.

\subsection{Further Analyses}
\label{sec.further_analyses}

\subsubsection{Discriminating Hard Negative Item Candidates.}
As stated, we aim to employ our model as a reranker in recommender systems. Previous experiments demonstrate its effectiveness in reranking a list of randomly sampled candidate items. In order to evaluate the model's ability in reranking more practical candidate items retrieved by a strong matching module, which are more challenging to distinguish compared to random negative items, we simulate the real \emph{matching-then-reranking} pipeline of recommendation. 

To this end, we introduce a matching module to retrieve a list of candidate items from all the items, then our model is instructed to rerank these candidates in the scenario of sequential recommendation. Specifically, we adopt the classical two-tower model as the matching module. The user tower incorporates the information of user ID, behavioral sequences' ID (encoded by a 2-layer transformer encoder), and behavioral sequences' titles (encoded by BERT). The item tower incorporates the information of item ID and item title (encoded by BERT). Therefore, the module considers both sequential patterns and textual similarities to retrieve candidate items. In this experiment, we retrieve nine negative candidate items with the positive target item. The results of our model and other baselines are reported in Table~\ref{tab:eval_deployment_with_recaller}. Our model still exhibits superior performance than other baselines with a notable gap when reranking the hard negative candidates. The result indicates that our model has a strong ability to discriminate and select items that are more in line with user information needs among similar items.

\begin{table}[!htbp]
\vspace{-1pt}
  \centering
  \caption{Performance on reranking hard negative candidates. We test it in scenario of sequential recommendation.}
    \resizebox{\columnwidth}{!}{\begin{tabular}{l|ccccc}
    \toprule
    \multirow{2.5}{*}{\textbf{Methods}} & \multicolumn{5}{c}{\textbf{Matching Module:} \bm{$\langle P_0, I_1, T_3 \rangle$}} \\
\cmidrule{2-6}          & HR@1  & HR@3  & HR@5  & NDCG@3 & NDCG@5 \\
    \midrule
    SASRec  & 0.0355 & 0.3448 & 0.5536 & 0.2127 & 0.2984 \\
    GPT-3.5 & 0.1100 & 0.3480 & 0.5020 & 0.2481 & 0.3113 \\
    \midrule
    InstructRec & \textbf{0.1841} & \textbf{0.4823} & \textbf{0.6648} & \textbf{0.3558} & \textbf{0.4307} \\
    Improv. & +67.36\% & +38.59\%    & +20.09\%    & +43.41\% & +38.36\% \\
    \bottomrule
    \end{tabular}}%
  \label{tab:eval_deployment_with_recaller}%
\end{table}%

\subsubsection{Discriminating More Candidate Items}
In previous experiments, we verify the effectiveness of our model in accommodating diverse interaction scenarios and hard negative samples by ranking a list of ten candidate items. However, in realistic recommender systems, it is commonplace to encounter a considerably larger pool of items (often numbering in the hundreds), which cater to user preferences. To evaluate the discriminative capabilities of our model across a broader range of candidate items, we simulate a rerank scenario for personalized search. This involved the random sampling of 99 negative items and the target item, resulting in a collection of 100 candidate items. Notably, as aforementioned, our model faces limitations in processing these candidate items concurrently due to the restricted context length. In order to complete this evaluation, we adopt a straightforward approach for prospective testing purposes. Specifically, we randomly divide the one hundred candidate items into ten equal groups and instruct the model to select the most potential item from each group. The resulting ten items would be then reranked, leading to the final ranking result of our model.

As demonstrated in Table~\ref{tab:eval_more_candidates}, our model exhibits a considerable performance advantage over the traditional baseline. We also find that the improvement observed in the HR@1 metric is relatively less significant compared to other metrics. This could potentially be attributed to the increased difficulty in distinguishing among the ten selected items. Consequently, besides employing LLM with longer maximum context lengths, the result also inspires us to explore efficient algorithms that can handle a larger number of candidates within the constraints of limited context length, which will be studied in future work.
Nevertheless, we believe that our InstructRec is more suitable for being deployed in the reranking stage (since it is the closest stage of recommender systems that users communicate with), and existing matching models are capable of retrieving good candidates in practice.

\begin{table}[htbp]
  \centering
  \caption{Performance on discriminating one hundred of candidate items. We test it in scenario of personalized search.}
  \begin{adjustbox}{width=1.0\columnwidth}
    \begin{tabular}{lccccc}
    \toprule
    \multirow{2.5}{*}{\textbf{Methods}} & \multicolumn{5}{c}{\bm{$\langle P_1, I_1, T_3 \rangle $}} \\
\cmidrule{2-6}          & HR@1  & HR@3  & HR@5  & NDCG@3 & NDCG@5 \\
    \midrule
    TEM   & 0.2794 & 0.4484 & 0.5284 & 0.3777 & 0.4106 \\
    InstructRec & \textbf{0.3276} & \textbf{0.7694} & \textbf{0.8334} & \textbf{0.5903} & \textbf{0.6163} \\
    Improv. & +17.25\% & +71.59\% & +57.72\% & +56.29\% & +50.10\% \\
    \bottomrule
    \end{tabular}%
      \end{adjustbox}
  \label{tab:eval_more_candidates}%
\end{table}%

\subsubsection{Effects of Instructions}
In this part, we explore the question that how the diversity of instructions affects the model's performance on held-out interaction scenario and evaluate its generalization. To do this, we set the scenario of personalized search with vague intention as a held-out scenario and continuously adding various types of fine-grained instructions for instruction tuning. Note that we refer to ``vague intention$^*$'' as another expression of user's vague intention simulated from the target review. That is, since review contains both user preferences and item characteristics. We employ the teacher-LLM to analyze user's vague intentions from the perspectives of both the item and the user, serving as training data for instruction tuning and testing data for evaluation.

\begin{figure}[!hbtp]
	\centering
	\includegraphics[width=0.7\linewidth]{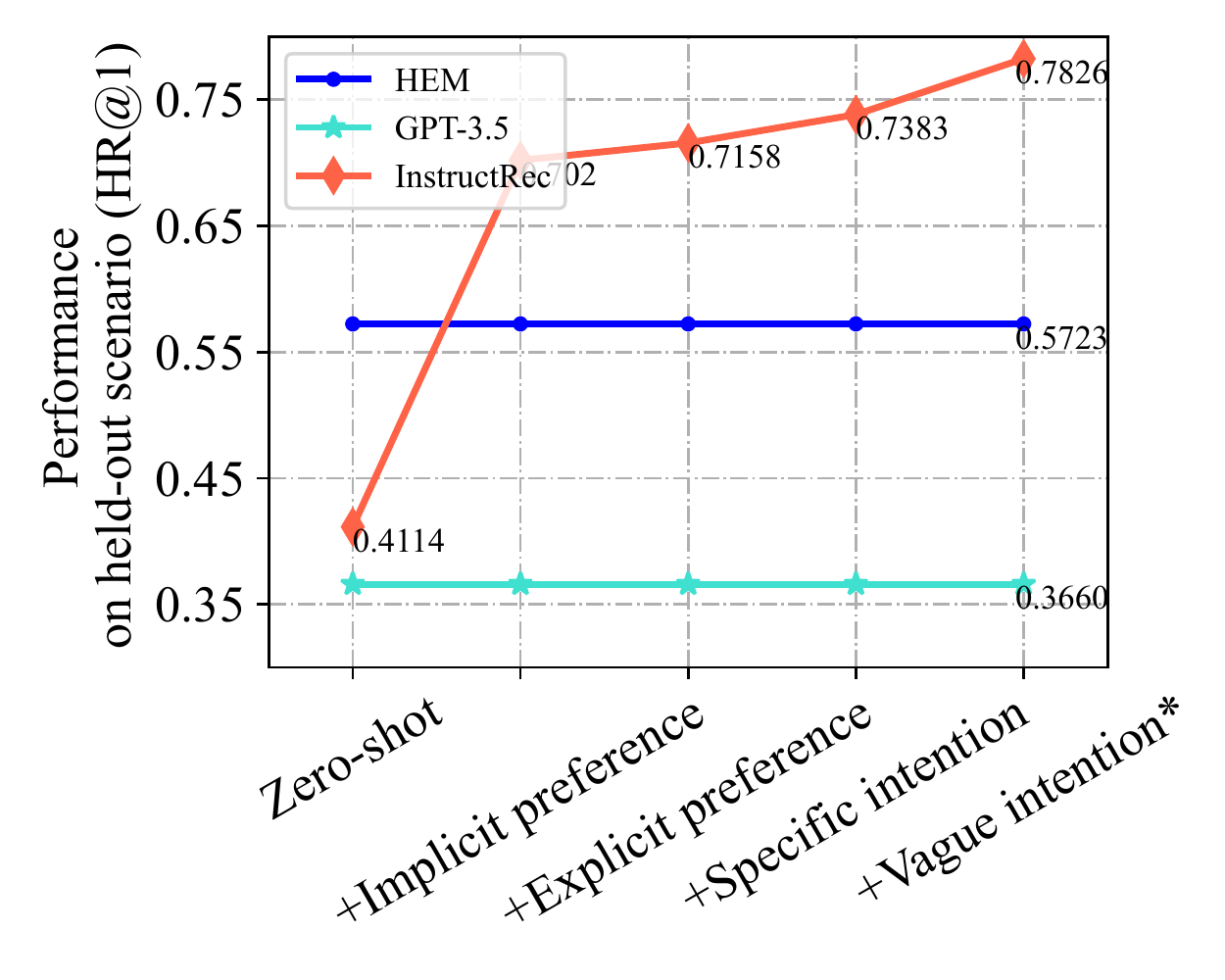}
	\caption{Instruction tuning on more interaction scenarios increases the performance on the held-out scenario. We take personalized search with vague intention as the held-out scenario.}
	\label{fig:eval_effect_of_instruction}
\end{figure}

As we can see in Figure~\ref{fig:eval_effect_of_instruction}, with the increasing number of interaction scenarios for instruction tuning, the performance of the model is steadily improved in the held-out interaction scenario. This observation demonstrates the effectiveness of instruction tuning in generalizing across various scenarios. Furthermore, it is worth noting that the incorporation of ``vague intention$^*$'' leads to a large improvement of performance, inspiring us to annotate more diverse data by conducting self-instruct with various strategies.

\begin{table}[htbp]
  \centering
  \caption{Performance on generalizing to unseen dataset. BERT4Rec and SASRec are trained on in-domain instances.}
  \begin{adjustbox}{width=1.0\columnwidth}
    \begin{tabular}{lccccc}
    \toprule
    \multirow{2.5}{*}{\textbf{Methods}} & \multicolumn{5}{c}{\bm{$\langle P_1, I_0, T_3 \rangle $}} \\
\cmidrule{2-6}          & HR@1  & HR@3  & HR@5  & NDCG@3 & NDCG@5 \\
    \midrule
    BERT4Rec & 0.5867 & 0.8226 & 0.9114 & 0.7249 & 0.7615 \\
    SASRec & \textbf{0.6713} & \textbf{0.8754} & \textbf{0.9428} & \textbf{0.7915} & \textbf{0.8194} \\
    \midrule
    GPT-3.5 & 0.1380 & 0.3140 & 0.4780 & 0.2401 & 0.3067 \\
    Flan-T5-XL & 0.0927 & 0.2886 & 0.4814 & 0.2035 & 0.2823 \\
    InstructRec$_{Games}$ & 0.2332 & 0.4392 & 0.6052 & 0.3511 & 0.4190 \\
    \bottomrule
    \end{tabular}%
    \end{adjustbox}
  \label{tab:eval_generalize_dataset}%
\end{table}%

\subsubsection{Generalization across Datasets}
Previous experiments have demonstrated the remarkable generalization capabilities of our model when it comes to accommodating unseen items, instructions and interaction scenarios. Despite the effectiveness, most of these evaluations focus on in-domain generalization. In this part, we aim to evaluate the model's ability to generalize to unseen datasets, which would have distinct patterns to the source data. To this end, we evaluate the model's ability to transfer from the ``Games'' dataset to the ``CDs'' dataset. The results are presented in Table~\ref{tab:eval_generalize_dataset}.

In general, our model's zero-shot performance in this setting still falls short compared to traditional sequential models that are trained on these in-domain instances. It is natural since in-domain behavioral information is essential in recommendation, which is implied in above evaluations. Nevertheless, Our model still outperforms other powerful LLMs that are good at zero-shot tasks by a large margin. It indicates that instruction tuning could bring in significant positive outcomes for our model, providing evidence that our approach can effectively capture universal knowledge across distinct domains.

\section{Conclusion and Future Work}
In this paper, we proposed an instruction tuning approach of LLMs for recommender systems, named \emph{InstructRec}. Different from existing studies that adapt LLMs for recommendation, our key idea is to consider recommendation as \emph{instruction following} by LLMs, allowing  users to freely express their information needs in natural language (called \emph{instructions}).  Specifically, we first designed the general instruction templates format by integrating the preference, intention, and task form, and context information of a user in natural language text. Then, we automatically generated 252K fine-grained user personalized instructions that describe user preferences and intentions. By tuning an open-source LLM (3B Flan-T5-XXL) with these instruction data, the base model can be well adapted  to recommender systems, which can follow user's instructions to perform personalized recommendation. Extensive experiments have demonstrated the effectiveness and generalization ability of our approach in various scenarios.

As future work, we will  further scale the size of LLMs for instruction tuning, and also consider extending the context length for modeling long behavior sequences.  
Besides, we will consider applying the current approach in a multi-turn interaction scenario, where users can communicate with the systems via a chit-chat way.

\bibliographystyle{ACM-Reference-Format}
\bibliography{reference}
\clearpage
\appendix
\section{Instruction Templates for \emph{Traditional Recommendation}} 
\noindent $\langle P_1, I_0, T_2\rangle$. Using the user's historical interactions as input data, predict the next product that the user is most likely to interact with. The historical interactions are provided as follows: \{\texttt{historical interactions}\}.

\noindent $\langle P_1,P_2, I_0, T_2\rangle$. Recommend the next potential product to a user based on his profile and past interaction. You have access to the user's profile information, including his preference: {{explicit preference}} and past interactions: \{\texttt{historical interactions}\}. Now you need to determine what product would be recommended to him.

\noindent \emph{Chain-of-thought (CoT)} like reasoning. Here is some information about the user, such as his historical interactions: \{\texttt{historical interactions}\}. Based on this information, your task is to infer the user's preference based on his historical interactions and recommend the next product to the user.

\noindent $\langle P_1, (P_2), I_0, T_2\rangle$. Given the following historical interaction of the user:  \{\texttt{historical interactions}\}. You can infer the user's preference. \{\texttt{explicit preference}\}. Please predict next possible item for the user.

\noindent $\langle P_1, (P_2), I_0, T_2\rangle$. Based on the historical interactions shown below: \{\texttt{historical interactions}\}, you can analyze the common ground among these interactions to determine the user's preferences \{\texttt{explicit preference}\}. Then please recommend a suitable item to the user.

\noindent $\langle P_1, (P_2), I_0, T_2\rangle$. To make a recommendation for this user, we need to analyze their historical interactions, which are shown below: \{\texttt{historical interactions}\}. As we know, historical interactions can reflect the user's preferences. Based on this user's preferences \{\texttt{explicit preference}\}, please recommend an item that you think would be suitable for them.

\noindent $\langle P_1, (P_2), I_0, T_3\rangle$. The behavioral sequence of the user is shown below: \{\texttt{historical interactions}\}, which can be used to infer the user's preferences \{\texttt{explicit preference}\}. Then please select the item that is likely to be interacted with the user from the following candidates, by comparing the candidates and their similarities to the user's preference. The candidates are: \{\texttt{candidate items}\}

\noindent $\langle P_1, (P_2), I_0, T_3\rangle$. You have observed that the user has clicked on the following items: \{\texttt{historical interactions}\}, indicating his personal tastes: \{\texttt{explicit preference}\} Based on this information, please select one item from the following options that you think would be suitable for the user: \{\texttt{candidate items}\}

\noindent $\langle P_1, (P_2), I_0, T_3\rangle$. You have some information about this user, which is shown below: \{\texttt{explicit preference}\}, the user's historical interactions: \{\texttt{historical interactions}\} Based on this information, please recommend the next possible item for the user, which should match the user's preference, from the following candidates: \{\texttt{candidate items}\}

\noindent $\langle P_1, (P_2), I_0, T_3\rangle$. You have obtained the user's historical interaction list, which is as follows:\{\texttt{historical interactions}\}. Based on this history, you can infer the user's preferences \{\texttt{explicit preference}\}. Now, you need to select the next product to recommend to the user. Please choose one from the following candidates: \{\texttt{candidate items}\}

\noindent $\langle P_1, (P_2), I_0, T_1\rangle$. The user has previously purchased the following items: \{\texttt{historical interactions}\}. This information indicates their personalized preferences \{\texttt{explicit preference}\}. Based on this information, is it likely that the user will interact with  \\ \{\texttt{candidate item}\} next?

\noindent $\langle P_1, (P_2), I_0, T_1\rangle$. Based on the user's historical interaction list, which is provided as follows: \{\texttt{historical interactions}\} , you can infer the user's personalized preference \{\texttt{explicit preference}\}. And your task is to use this information to predict whether the user will click on \{\texttt{candidate item}\} next.

\noindent $\langle P_1, (P_2), I_0, T_3\rangle$. Please recommend an item to the user based on the following information about the user: \{\texttt{historical interactions}\} ,the user's historical interaction, which is as follows: \{\texttt{explicit preference}\}  Try to select one item from the following candidates that is consistent with the user's preference: \{\texttt{candidate item}\}.

\noindent \emph{Turn the task around}. You have the ability to infer a user's preferences based on his past interactions. You are provided with a list of the user's past interactions : \{\texttt{historical interactions}\}  Your task is to analyze the commonalities among the past interactions and infer his overall preference. Please provide your analysis and inference.

\noindent \emph{Turn the task around}. As we all know, the user's historical interactions are guided by his personalized preference. Try to infer the user's preferences by analyzing his historical interactions : \{\texttt{historical interactions}\}

\noindent $\langle P_2, I_0, T_2\rangle$. You are a recommender system, and are good at recommending products to a user based on his preferences. Given the user's preferences: \{\texttt{explicit preference}\}, please recommend products that are consistent with those preferences. 

\noindent $\langle P_2, I_0, T_2\rangle$. As we know, a user's behavior is driven by his preferences, which determine what they are likely to buy next. Your task is to predict what products a user will purchase next, based on his preferences. Given the user's preferences as follows: \{\texttt{explicit preference}\}, please make your prediction.

\section{Instruction Templates for \emph{Traditional Product search}} 

\noindent $\langle P_0, P_2/I_1/I_2, T_2\rangle$. Suppose you are a search engine, now the user search that \{\texttt{explicit preference}/ \texttt{vague intention}/ \texttt{specific intention}\}, can you generate the item to respond to user's query?

\noindent $\langle P_0, I_2, T_2\rangle$. As a search engine, your task is to answer the user's query by generating a related item. The user's query is provided as \{\texttt{specific intention}\}. Please provide your generated item as your answer.

\noindent $\langle P_0, I_2, T_2\rangle$. As a recommender system, your task is to recommend an item that is related to the user's request, which is specified as follows: \{\texttt{specific intention}\} Please provide your recommendation.

\noindent $\langle P_0, P_2/I_1/I_2, T_2\rangle$. \quad If a user asks a question like: \\ \{\texttt{explicit preference}/ \texttt{vague intention}/ \texttt{specific intention}\} Please generate a related answer to help him.

\noindent \emph{Turn the task around}. If a user wants to search for something specific in a search engine but doesn't know how to phrase the query, we can help generate the query for them. Now the user wants to search for \{\texttt{target item}\}. Please generate the query. 

\noindent \emph{Turn the task around}. As a search engine that has seen many user queries, you can make an educated guess about how a user might write a query when searching for a particular item. If a user were searching for the item: \{\texttt{target item}\} They might use keywords related to the item such as its brand, or type. So the query would be.

\noindent $\langle P_0, P_2/I_1/I_2, T_3\rangle$. You are a search engine and you meet a user's query \{\texttt{explicit preference}/ \texttt{vague intention}/ \texttt{specific \\ intention}\}. Please respond to this user by selecting items from the candidates:  \{\texttt{candidate items}\}

\noindent $\langle P_0, P_2/I_1/I_2, T_3\rangle$. Your task is to select the best item from a list of candidates that meets the user's needs based on their search query. Here is the search query of the user: \texttt{explicit preference}/ \texttt{vague intention}/ \texttt{specific intention}\} and the candidates are as follows: \{\texttt{candidate items}\}

\noindent $\langle P_0, P_2/I_1/I_2, T_3\rangle$.  Your task is to select the best item from a list of candidates that matches the user's query, by comparing the candidate list and their relevance to the user's query. The user has entered the following search query: \texttt{explicit preference}/ \texttt{vague intention}/ \texttt{specific intention}\} And here are the candidate list: \{\texttt{candidate items}\}

\section{Instruction Templates for \emph{Personalized Search}} 

\noindent $\langle P_1, P_2, T_2\rangle$.  You are a search engine. Here is the historical interaction of a user: \{\texttt{historical interactions}\}. And his personalized preferences are as follows: \{\texttt{explicit preference}\}. Your task is to generate a new product that are consistent with the user's preference.

\noindent $\langle P_1, I_1, T_2\rangle$.  The user has interacted with a list of items, which are as follows: \{\texttt{historical interactions}\}. Based on these interacted items, the user current intent are as follows \{\texttt{vague intention}\}, and your task is to generate products that match the user's current intent.

\noindent $\langle P_1, I_1, T_2\rangle$.  As a shopping guide, you are assisting a user who has recently purchased the following items: \{\texttt{historical interactions}\} The user has expressed a desire for additional products with the following characteristics: \{\texttt{vague intention}\} Please provide recommendations for products that meet these criteria.

\noindent $\langle P_1, I_2, T_2\rangle$.  As a search engine, you are assisting a user who is searching for the query: \{\texttt{specific intention}\}. Your task is to recommend products that match the user's query and also align with their preferences based on their historical interactions, which are reflected in the following: \{\texttt{historical interactions}\}

\noindent \emph{Turn the task around}.  The user has recently purchased the following items: \{\texttt{historical interactions}\} Now he is interested in finding information about an item that he believe he still need, which is: \{\texttt{target item}\}. However, the user is unsure how to write a query to search for this item based on their preferences. Please assist the user in writing a query.

\noindent \emph{Turn the task around}.  The user has the following historical interactions: \{\texttt{historical interactions}\}. And he is interested in purchasing the target item: \{\texttt{target item}\}. Please analyze the user's historical interactions and identify his preferences that lead the user to interact with the target item."

\noindent \emph{Enforcing the relatedness between preference and intentions}.  The user has searched for the query:  \{\texttt{explicit preference}/ \texttt{vague intention}\} and ultimately selected the product:  \{\texttt{target item}\} Based on the user's query and final choice, you can infer their preferences. Additionally, the user's historical interactions have also been influenced by their preferences. Please estimate the user's historical interactions that match their preferences, taking into account their search query and final product selection.

\noindent \emph{Enforcing the relatedness between preference and intentions}.  As a search engine, you have observed the following behavioral sequence of a user: \{\texttt{historical interactions}/ Using the content and categories of the user's historical interactions, you can infer their preferences. Please make a prediction about the user's next query and the product he is likely to ultimately purchase, based on his preference.

\noindent \emph{Turn the task around}.  A user's query can provide insight into their preferences as well as their future purchasing intentions. Furthermore, a user's behavior is often influenced by their preferences. Given the user's query: \{\texttt{explicit preference}/ \\ \texttt{vague intention}\} Please analyze the query to speculate on what products the user has previously purchased and predict what products they are likely to purchase next based on their past queries and preferences.

\noindent  $\langle P_1, (P_2), P_2/I_1/I_2, T_3\rangle$.  Using the user's current query: \{\texttt{explicit preference}/ \texttt{vague intention} / \texttt{specific intention}\} and their historical interactions: \{\texttt{historical interactions}\} you can estimate the user's preferences \{\texttt{explicit preference}\} Please respond to the user's query by selecting an item from the following candidates that best matches their preference and query: \\ \{\texttt{candidate items}\}

\noindent  $\langle P_1, (P_2), I_1/I_2, T_3\rangle$.  The user wants to buy some products and \\ searches for: \{\texttt{explicit preference}/ \texttt{vague intention} / \\ \texttt{specific intention}\}. In addition, they have previously bought:  \{\texttt{historical interactions}\} You can estimate their preference by analyzing his historical interactions.  \{\texttt{explicit preference}\} Please recommend one of the candidate items below that best matches their search query and preferences:  \{\texttt{candidate items}\}

\noindent  $\langle P_1, (P_2), I_1/I_2, T_3\rangle$.  A user enjoys shopping very much and has purchased a lot of goods. They are: \{\texttt{historical interactions}\}. His historical interactions can reflect his personalized preference. \{\texttt{explicit preference}\}. Now he wants to buy some new items, such as: '\{\texttt{explicit preference}/ \texttt{vague intention} / \texttt{specific intention}\}' The recommender system recommends the following candidate items based on his needs and preferences: \{\texttt{candidate items}\} Please select the item that best meets the user's needs and preferences from among the candidate items.

\noindent  $\langle P_1, (P_2), I_1/I_2, T_3\rangle$.  Based on the user's historical interactions with the following items: \{\texttt{historical interactions}\} You can infer his preference by analyzing the historical interactions. \{\texttt{explicit preference}\} Now the user wants to buy a new item and searches for: ``\{\texttt{explicit preference}/ \texttt{vague intention} / \\ \texttt{specific intention}\}'' Please select a suitable item from the following candidates that matches his preference and search intent: \{\texttt{candidate items}\}.

\end{document}